\documentstyle[twocolumn,prl,aps,psfig]{revtex}
\tightenlines
\begin{document}
\draft
\title{Ground state and far-infrared absorption of 
two-electron rings in a magnetic field}
\author{Antonio Puente and Lloren\c{c} Serra}
\address{Departament de F\'{\i}sica, Universitat de les Illes Balears,
E-07071 Palma de Mallorca, Spain}
\date{May 30, 2000}
\maketitle
\begin{abstract}
Motivated by recent experiments  
[A. Lorke {\em et al.}, Phys.\ Rev.\ Lett.\ {\bf 84}, 2223 (2000)]
an analysis of the ground state and far-infrared absorption
of two electrons confined in a quantum ring is presented. The height 
of the repulsive central barrier in the confining potential is shown to 
influence in an important way the ring properties. 
The experiments are best explained assuming the presence of both 
high- and low-barrier quantum rings in the sample.
The formation of a two-electron Wigner molecule in certain cases is 
discussed.
\end{abstract}

\section{Introduction}

The possibility of confining a controlled number of electrons in 
artificial nanostructures at the interface of two semiconductors, 
achieved in recent years\cite{books}, has attracted much interest.
An obvious reason is the potential application of these systems 
to sub-micron electronics technology but, from the theoretical 
point of view, much importance is attributed to their character
of novel quantum systems in which new physics, as well as analogies 
with atomic clusters, atoms or nuclei, may be found. The most widely studied 
systems have been the so-called 'quantum dots', in which a certain 
number of electrons form a two-dimensional compact island, modelled by 
a confinig potential of parabolic type (for small dots) plus the mutual
electron-electron interactions. The electronic islands have been also 
formed in the shape of rings. In fact, mesoscopic rings were produced 
by Dahl {\em et al.}  and their magnetoplasmon resonances measured for different 
ring widths \cite{Dah93}. 
Subsequent theoretical models, using semiclassical 
methods \cite{Zar96} and density functional theory \cite{Emp99} 
provided a good interpretation of the measured ring spectra.
There also exist approaches based on exact diagonalization methods 
for very small numbers of electrons which have addressed the so called 
'persistent current'
\cite{Tap94}
and the optical excitations for very narrow 
rings (approaching the 1d limit) as well as for dots with a repulsive 
scatterer center \cite{Hal96,Niem96}. Besides, a description of the 
optical properties in two electron rings based on an analytical 
treatment of the ground state was given by Wendler {\em et al.}
\cite{Wen96}. 

Very recently, Lorke and co-workers  reported the fabrication, using 
self-assembly techniques, of electron rings in the true quantum limit, 
not affected by random scatterers, containig two electrons \cite{Lor00}. 
The far-infrared (FIR)
spectra as a function of the magnetic field has shown the appearance of
electronic excitations peculiar of a ring geometry as well as a 
magnetic-field induced transition in the ground state angular momentum.
The two electron system is the smallest one with electron-electron interaction
and its reduced size places it within the reach of essentially exact methods, 
such as the Hamiltonian diagonalization in a basis of functions mentioned 
above. In this paper we have followed an alternative although also exact 
method, based on the solution of the Schr\"odinger equation by discretizing 
the coordinate space in a uniform grid of points. 
The method is therefore free from basis selection and truncation although its 
limitation lies, obviously, in the number of spatial points. This scheme can 
also be used to obtain the dynamical properties by solving the 
time-dependent Schr\"odinger equation and monitoring the evolution of  
relevant physical quantities with time. In this work we have addressed 
the linear response regime, for small amplitude oscillations. However, one of 
the potentialities of the real-time approach is the study of non-linear 
(large amplitude) dynamical processes.
By finding the exact solution one can easily quantify the 
effect of the electron-electron interaction as well as 
check the validity of approximate descriptions, such as 
density-functional theory. In fact, an analysis
of the experiment based on the local spin-density approximation (LSDA) in a 
symmetry-restricted approach has already been presented in Ref.\ \cite{Emp99b}. 

An important input to the theory is the external confining potential 
acting on the electrons. Since we aim at a direct comparison with experiment,
the strategy should be to introduce a reasonable model potential,
depending only on a few parameters and explore the exact solution 
as a function of this parameters. Our approach has been to take the parabola 
parameter $\omega_0$ and ring radius $R_0$ as guessed in Ref.\ \cite{Lor00} 
and explore the effect of the central repulsive barrier by solving for a low-,
an intermediate- and a high-barrier ring. With these barriers the system 
changes from a compact dot-like structure to a well developed 
ring one.  We show that the barrier height 
modifies the ring properties and that the experiments are best explained 
assuming a mixed sample containing both low- and high-barrier rings.

In recent papers Yannouleas and Landman \cite{Yan00}, 
as well as Koskinen {\em et al.} \cite{Kos00}
have discussed the electronic localization, in the relative frame,
known as the formation of {\em Wigner molecules}
in two electron quantum dots
and in rings with 6 electrons, respectively.
Both groups have addressed the problem by calculating  
the low-lying states and discussing their rotor-like structure. This 
crystallization is in good agreement with the prediction of mean-field (MF) 
theories like the Hartree-Fock or the Local-Density-Functional approaches. 
We also comment on the formation of a Wigner molecule in the high-barrier 
rings by comparing the conditional probability distribution with the LSDA 
density. In this case, the LSDA predicts a symmetry-broken ground state that 
nicely agrees with the electronic distribution built from the conditional 
probability density of the exact solution to the Schr\"odinger equation. 

The structure of the paper is as follows. Section II is devoted to the 
exact solution for the ground state and Sec.\ III to the corresponding 
exact solution for the dynamics. In Sec.\ IV we discuss the comparison with 
the mean field approach and the formation of the Wigner molecule.
Finally, the conclusions are presented in Sec.\ V.
   
\section{Ground state}

\subsection{Hamiltonian and resolution method}

The Hamiltonian of the two-electron system, in a magnetic field,
using effective units \cite{units}, is 
\begin{eqnarray}
\label{eq1}
H&=&\sum_{i=1,2}\left[
{1\over2}
\left( -{\rm i}\nabla_i+\gamma {\bf A}({\bf r}_i) \right)^2
+V({\bf r}_i)\right] \nonumber\\
&+& {1\over |{\bf r}_1-{\bf r}_2|}
+g^* m^*{\gamma\over 2}\,{\bf B}\cdot{\bf S}
\; ,
\end{eqnarray}
where ${\bf A}$ is the vector potential, $\gamma=e/c$ (we assume Gaussian
magnetic fields) and $V({\bf r})$ is the confining external potential.
The electronic positions are restricted to the $xy$ plane, i.e., 
${\bf r}\equiv(x,y)$, and we consider a constant and perpendicular magnetic 
field ${\bf B}=B {\bf e}_z$, 
described in the symmetric gauge with a vector potential 
${\bf A}({\bf r})\equiv B/2(-y,x)$. The last piece is the Zeeman term,
depending on the total spin ${\bf S}={\bf s}_1+{\bf s}_2$ 
the effective gyromagnetic factor $g^*$ and electronic effective mass 
$m^*$\cite{units2}.

The ring external potential is modelled by the following circularly symmetric 
piecewise function,
\begin{eqnarray}
\label{eq2}
V(r)=\left\{
\begin{array}{ll}
{1\over 2} \omega_0^2 (r-R_0)^2 & \quad {\rm if}\; r\geq R_0 \\[0.1cm]
{1\over 2} \omega_0^2 (r-R_0)^2\, [1 + & \\
\quad c (r-R_0) + d (r-R_0)^2] & \quad {\rm if}\; r < R_0
\end{array}\right.
\; .
\end{eqnarray}
In this expression $R_0$ is the ring radius and $\omega_0$ gives the curvature 
around the minimum at $R_0$. 
Fixing a barrier height $V_b$ at the origin determines 
the parameters $c$ and $d$, with the additional requirement of 
smootheness at $R_0$ \cite{note}.

Since Hamiltonian (\ref{eq1}) depends on spin only through the Zeeman term, 
the spin part of the wave function is given by the well known singlet 
and triplet combination of spinors.  
Our method to obtain the spatial wave function is based on the only assumption 
that 
its gradient and Laplacian can be obtained by using 
finite difference formulas in a spatial grid of uniformly spaced points. 
Associating a macroindex $I\equiv(x_1,y_1,x_2,y_2)$ to each point and 
dicretizing the derivative operators (we use 5 point formulas), 
the Schr\"odinger eigenvalue equation transforms into the linear problem
\begin{equation}
\label{eq4}
H_{IJ}\Psi_{J}=E\Psi_I\; .
\end{equation} 
It is worth to mention that the interaction and potential terms are diagonal
(local) and only the kinetic contribution gives 
non diagonal elements. Besides, since the derivatives involve only 
a finite number of points (we have tipically used 5-point formulas)
the Hamiltonian matrix is very sparse. We have solved Eq.\ (\ref{eq4}) for 
the ground state using the iterative 
{\em imaginary time-step} method \cite{Koo90}.
In principle, one can also obtain excited states by introducing the 
constraint  of orthogonalization to the lower eigenvectors, although the 
computational effort increases very rapidly. A very useful optimization is 
introduced by considering only symmetric (singlet) or antisymmetric (triplet) 
wave functions with respect to the exchange of the two particle positions, 
since this allows to reduce by a factor two the wave function dimension. 

The above method has proved to be quite stable and the 
convergence with the number of points very fast. Typically, we have used 
from $N_1\approx$20 up to $\approx$50 points for the discretization of a single 
dimension. Notice that the wave function dimension ($N_1^4$) 
would normally imply that the Hamiltonian matrix 
($N_1^4\times N_1^4$) exceed the storage capability. This is avoided 
by fully exploiting the sparseness of the Hamiltonian and not really 
dimensioning such a big matrix. Analogously, the imaginary-time step 
method allows us to determine the lower eigenstate of this matrix
which would be unfeasible by direct diagonalization methods. 
We have also checked the convergence with the number of points used 
to discretize the Laplacian operator, {\em e.g.}, when going from 5 
to 9 point formulas a relative difference less than $1/10^3$ in the 
ground state energy is found.

\subsection{Results}

In this subsection we present the results for the ground state of  
rings having $R_0=1.4$~$a_0^*$ and 
$\omega_0=1.1$~${\rm H}^*$. The corresponding physical values are, 
approximately, 14 nm and 13 meV and are taken as reasonable values from
the experiment. Three different barrier heights are considered $V_b=0.75$, 
3 and 12 H$^*$. As shown in Fig.\ 1 these three values correspond, going from 
low to high barrier, to decreasing central densities. 
In what follows the three rings will be referred to as V1, V2 and V3 
for increasing barrier height, respectively.
While ring V1 has a compact density because of its rather low barrier, 
V3 has a real ring shape, with a hollow central region. 

Figure 2 displays the expectation value of the angular momentum 
$\langle L_z\rangle$ as well as the energy of the lowest level 
for both singlet and triplet states as a function of the magnetic field.
At $B=0$ the ground state is a singlet and with increasing $B$ there is 
an alternation of triplet and singlet ground states in the three rings.
The first singlet-triplet transition occurs approximately at 
3, 2 and 1 T for V1, V2 and V3, respectively. 
The triplet-singlet energy differences remain quite small after the 
first crossing which hints at possible transitions induced by 
small non-spin conserving perturbations like spin-orbit
interactions or magnetic impurities.
We also notice how the ground state increases in steps its angular momentum 
(in absolute value) as a function of $B$. This reflects the exact 
property that angular momentum is a good quantum number since the Hamiltonian 
commutes with the $L_z$ operator.
The angular momentum gain with magnetic field is faster for the higher barrier 
rings. In fact, for V3 the evolution is markedly linear,
already indicating a rotor-like evolution for this ring. This is further 
discussed in Sec.\ IV. 

\subsection{Non-interacting solution}

When neglecting the Coulomb interaction the Hamiltonian (\ref{eq1})
separates for each particle and, therefore, a solution built from 
single-particle (sp) orbitals with radial $n$ and angular $\ell$ 
quantum numbers is adequate\cite{books}. 
Figure 3 shows the sp eigenvalues for all, low and high barrier, rings. 
In fact, from the lowest maximum of the capacitance-voltage
spectrum Lorke {\em et al.} obtained the single-electron ground-state
shift with magnetic field. This manifests a change in slope at
$B\sim 8$T that was explained as the transition from $\ell=0$ to $\ell=-1$
in a model potential.
Of course, this is in agreement with the sp levels for V1, since the 
low-barrier ring essentially coincides with the model potential used 
in Ref.\ \cite{Lor00}. The first crossing for V2 is located around $4$T, 
 while as shown in Fig.\ 3 the high barrier 
case (V3) has again a change in slope at $B\sim 8$T, attributed to the 
$\ell=-1\longrightarrow\ell=-2$ transition, 
which may indicate that the 
experimental sample could contain both low and high barrier rings.
This point will be further discussed when presenting the FIR
spectra.  

The non-interacting ground state with two electrons is obtained by filling the
lowest orbitals of the sp scheme, with the constraint of having a singlet 
or triplet spin. 
The corresponding plots for the evolution of $\langle L_z\rangle$
and the ground state energy as a function of magnetic field are displayed in 
Fig.\ 4. By comparing it with Fig.\ 2 we notice the effects of the 
Coulomb interaction between the two electrons. Obviously, the non-interacting 
case gives lower ground state energies since the Coulomb repulsion is neglected. 
Due to the increasing effect of the confining potential the difference is 
higher for V1 than for V3. Differences in angular momentum are also higher for 
the low barrier ring. This is indicating that in well developed rings the 
interaction effects are less important than in more compact structures, 
in agreement with the findings of Ref.\ \cite{Tap94} and with the obvious 
expectation of a larger $V_b$ keeping the electrons more apart 
from each other than a lower barrier.

We also realize that the triplet solution is unfavoured in the non-interacting 
case. In V1 the gs has always $S=0$, except 
for $B=8$ T for which singlet and triplet are essentially degenerate.
A similar thing happens in the other rings, with the difference that 
the singlet-triplet degeneracy points appear at lower $B$'s. 
These differences with respect to the exact results, in which clearer
singlet-triplet oscillations are seen, indicate 
the importance of the Coulomb interaction for a proper ground state 
description.

\section{Time integration}

The time evolution of the two-electron wave function is given by the 
time-dependent Schr\"odinger equation
\begin{equation}
\label{eq5}
{\rm i} {\partial\over\partial t}\,\Psi({\bf r}_1,{\bf r}_2;t)=
H \Psi({\bf r}_1,{\bf r}_2;t)\; ,
\end{equation}
where $H$ is the Hamiltonian operator (\ref{eq1}). We have solved 
Eq.\ ({\ref{eq5}), in the same spirit as for the ground state, by 
discretizing the coordinate space in a uniform grid of points. Time is also 
discretized and the evolution from time $(n)$ to $(n+1)$ is obtained through 
the unitary Crank-Nicholson algorithm 
\begin{equation}
\label{eq6}
\Psi^{(n+1)}_{I}=\Psi^{(n)}_{I}-
{\rm i} {\Delta{t}\over 2} 
\sum_J{H_{IJ}\left(\Psi^{(n)}_{J}+\Psi^{(n+1)}_{J}\right)} \; .
\end{equation}
This is an implicit scheme, in which the new $\Psi^{(n+1)}$ may be obtained 
by iteration. The above algorithm allows to perform the dynamical simulation of 
Eq.\ ({\ref{eq5}) by using a small time step $\Delta{t}$.
Of course, the feasibility of this approach to the solution of 
Eq.\ (\ref{eq5}) relies on the very small number of constituents 
(two electrons) of the system.

In this paper we discuss the FIR absorption of two-electron rings
and therefore our interest will focuss on the dipole states corresponding to 
small amplitude charge oscillations. The ground state wave function is 
perturbed by an initial (small) rigid displacement in a certain 
direction $\hat{\bf e}$, representing the electric field direction,
and the dipole moment evolution is then followed in time. 
The absorption cross section is given by the frequency transform of the 
dipole signal
\begin{equation}
\label{eq7}
\sigma(\omega) \approx
{\omega\over 2\pi} \left\vert \int{dt  
\langle \Psi(t) | D | \Psi(t) \rangle }\,
\exp(i \omega t) \right\vert \;,
\end{equation}
where the dipole operator is 
$D=\hat{\bf e}\cdot({\bf r}_1+{\bf r}_2)$.
As a test of the method, developed from previous calculations
within mean-field approaches, we have checked that the oscillation 
frequencies when assuming a pure parabolic potential are given, as a 
consequence of the generalized Kohn theorem, by the values \cite{Mak90}
\begin{equation}
\label{eq7b}
\omega_\pm=\sqrt{\omega_0^2+\omega_c^2/4}\pm{\omega_c/2} \; ,
\end{equation} 
where $\omega_c=eB/c$ is the cyclotron frequency.

\subsection{FIR spectra}

Figure 5 shows the absorption spectra
for singlet and triplet states corresponding to different magnetic
fields. The intersection of the baseline for each spectrum with the 
horizontal axis indicates the magnetic field in each case.
Notice that the horizontal scale for each spectrum is arbitrary, and
therefore, it is not given by the horizontal axis.
The spectra are drawn in linear scale and arbitrary units while  
the vertical scale gives the energies in meV. Not surprisingly the 
V1 spectrum is very similar to 
that of compact dots, with two branches of opposite $B$-dispersion which 
merge at $B=0$. This is indeed evident by comparing the spectra 
with the analytical expressions (\ref{eq7b}) (dashed lines), where 
an empirical $\omega_0$ value fixed to the lowest $B=0$ peak energy  
has been used. 
Except for some fragmentation details these two branches are rather
similar for both singlet and triplet cases. More importantly, we want to 
stress that essentially no strength is contained in the 
energy region $15\; {\rm meV} < \omega < 25\; {\rm meV}$ for $B<6$~T.
This fact does not imply that there are no excitations in this region.
In fact they can be seen using logarithmic scale,
as has been done in the LSDA approach of Ref.\ \cite{Emp99b}. 
We have sticked however to the linear scale firstly because the real-time 
approach requires extremely long simulation times in order to extract so low 
intense frequencies, and secondly we do not think that the experiments can  
easily identify such low intensity peaks.

The V2 spectrum and, more clearly, the V3 one reflect a behaviour
qualitatively different from that of V1. The high energy branch 
does not merge with the low one for vanishing $B$, deviating from the 
analytical prediction of Eq.\ (\ref{eq7b}). This behaviour 
is well known for quantum rings and is a manifestation of the 
violation of Kohn's theorem in these systems 
\cite{Dah93,Zar96,Emp99,Tap94,Hal96,Niem96,Wen96}.
For V2 a global gap $\approx 8-14$ meV persists at all magnetic fields 
for both spins,  while in the case of V3 the corresponding gap lies 
approximately in the region 
$7-17$ meV. Therefore, this gap constitutes a direct manifestation of the 
barrier height in the FIR response. The $B$-evolution of the spectra is 
characterized by small jumps in the peak positions for each branch 
which must be attributed to changes in total angular momentum. 
Since V3 is the one gaining more angular momentum its spectrum also 
displays more abundant discontinuities. We also notice that when the 
ring barrier 
grows (from V2 to V3) an important amount of strength is placed 
in the interval 15-25 meV at low magnetic fields. Actually the V3 peak 
in this region has a strength comparable to the one at low energy 
$\approx 50\%$, as found experimentally \cite{Emp99b}, while our 
calculation for V1 as well as the LSDA results of Ref.\ \cite{Emp99b} 
yield a much lower ratio.

A comparison with experiment of the calculated spectra reproduces 
in each single case some of the experimental features but misses others. 
A plausible explanation is that the sample contains a mixture of high and low 
barrier rings. In Fig.\ 6 we display a superposition of the V1 and V3 spectra 
in comparison with the available experimental points, assuming that the system 
takes the theoretical ground state at each magnetic field, although as 
already noticed in many cases the triplet-singlet energy difference is 
quite small.
With this interpretation, the circles are attributed to high barrier 
rings while the triangles and rombuses would correspond to low-barrier 
ones. Although the positions of the experimental peaks are not perfectly 
reproduced (this would involve a further optimization with respect to 
$\omega_0$ and $R_0$) the qualitative agreement is good. 
 A richer mixture including intermediate-barrier rings would induce a 
blurring of the different branches with the introduction of additional 
(intermediate) peaks, but would also yield the above discussion and 
comparison with experiment more involved. The crosses in Fig.\ 6 are 
not reproduced by any of our ring calculations and, as hinted by the 
experimental group, they could be due to pure quantum dots in the 
sample. We also mention that in the experiment \cite{Lor00} an 
insufficient signal to noise ratio at low energies does not allow the 
detection of the lower branch at $\omega < 10 \; {\rm meV}$.

\subsection{Non-interacting spectra}

Figure 7 shows the singlet and triplet spectra when the Coulomb interaction 
is neglected as in subsection II.C. Therefore, they correspond to pure 
particle-hole transitions in the single-particle level scheme. On a first
look they seem quite similar to the spectra of Fig.\ 5. However, after 
a second inspection important differences appear. A dramatic one 
is the softening, i.e., a decrease in energy, of the lowest branch 
in all three cases. In fact for 
$B\approx 8$ T the lowest state has an almost vanishing energy. This fact 
is explained as the formation of quasi Landau bands, even for such a small 
system, with the property that single-particle states of the same band are 
quasi-degenerate and therefore, particle-hole transitions are nearly gapless.

There are also sizeable shifts of peak positions and, in general, a higher 
fragmentation is present in the single particle picture. This is easy 
to understand since the collectivity asociated to the Coulomb interaction
is absent in the second case, and therefore the description of 
collective magnetoplasmon states is quite rough. 
Altogether, the correct 
qualitative result must be understood as an indication that interaction 
effects are not overwhelming for the spectrocopic features of a 
two-electron system.  

\section{Mean field theory and broken-symmetry solution}

Usually, the exact solution of the many-body Schr\"odinger equation
is not possible and one must content with approximate solutions.
In this sense, the mean field aproximation allows in many cases 
the description of relevant physics. Although in principle 
the mean field provides an accurate picture for a big enough 
number of particles, it is interesting to compare it with the 
exact solution in the present case for two particles. This will 
provide a strong test of the mean field result, that 
will allow us to quantify its validity. We remark
that for systems confined in an external potential such as 
atomic or quantum dot electrons the restriction on having a sufficient number
of particles in order to develop the mean field is not as strict 
as in self-bound systems, such as liquid drops or atomic nuclei.  
This is easy to understand since in confined systems the fixed external 
potential is already an important contribution to the mean field
and, therefore it is not solely based on particle-particle 
interactions.

Imposing circular symetry the mean field theory has indeed been applied to 
medium-large quantum dots ($N\approx 20-200$), in the Hartree \cite{Kemxx}, 
Hartree-Fock \cite{Gudxx} and LSDA schemes \cite{Pixx,Reixx}.
Nevertheless when using symmetry-unrestricted approaches, even at the mean 
field level, it is normally necessary to restrict to smaller sizes ($N\sim 10$)
because of the computational requirements \cite{Yan99,Pue99,Ull00,Gud00}.  

\subsection{LSDA ground state}

We have applied the LSDA symmetry-unrestricted approach developed by us in
Ref.\ \cite{Pue99} to the present two-electron rings. Figure 8
summarizes the ground state properties, for both singlet and triplet 
for V1, V2 and V3. The comparison with the 
energies for the exact solution (Fig.\ 2) prove that the the mean field 
is providing a quantitatively good description of the ground state energy, 
with relative energy differences around 2\%.
We notice that the MF description of the triplet is slightly better than  
that of the singlet. In fact, after the first singlet-triplet crossing the 
MF fails to reproduce a singlet state more bound than the triplet one.
This can be attributed to a more important effect of correlations due 
to the interaction in the singlet than in the triplet which are only 
approximately taken into account within LSDA. 
The overall evolution of the total 
angular momentum is also well reproduced in MF, especially for the 
triplet. Quite strikingly, however, the angular momentum of the 
singlet does not show the step-like evolution peculiar of the 
exact solution. In fact it is taking in most cases clear non-integer 
values. This must be attributed to a spontaneous breaking of the 
rotational symmetry in the frame of the mean field. 

The symmetry of the MF solution is more easily appreciated in Fig.\ 9,
that shows the MF ground state density 
$\rho=\rho_\uparrow+\rho_\downarrow$ and magnetization 
$m=\rho_\uparrow-\rho_\downarrow$, where $\rho_\uparrow$, 
$\rho_\downarrow$ are the two spin densities,
for V1 and V3 at $B=0$. Both cases correspond to singlet solutions
but, while V1 has a 
circularly symmetric density and vanishing magnetization, V3 is showing a
separation of spin up and down densities, leading to an 
oscillation of the magnetization along the ring perimeter 
(a spin density wave) as well as of the total density (a charge density wave).
 This behaviour is in fact common to all singlet states at $B>0$ 
for which $\langle L_z \rangle$ takes non-integer values. 
This peculiar states were predicted by Reimann {\em et al.} \cite{Rei99} 
in narrow rings with a bigger number of electrons, and have been 
also studied by us in Ref.\ \cite{Val00}. 

The formation of broken-symmetry solutions within MF approaches 
is a well known phenomenon in nuclear physics \cite{Rinxx}.  
In this field, it has been shown that the MF is in fact providing the system 
structure in an {\em intrinsic reference frame}, given by 
proper {\em intrinsic} coordinates, while there is 
degeneracy with respect to the remaining {\em collective} coordinates.
In a simplifying picture this corresponds to assume a certain structure of 
the system relative to its mean field, that in turn may be moving thus 
leading to a restoration of the exact symmetry. In fact, as shown in Fig.\ 1, 
when performing an angular average of the MF density for V3 an excellent  
agreement with the exact result is obtained. As we discuss further in the next 
subsection, the MF symmetry breaking can be interpreted as an electronic 
crystallization, i.e., the formation of a Wigner molecule.

\subsection{The Wigner molecule}

The symmetry breaking of the MF solution for V3 is pointing an 
incipient electron localization in the intrinsic frame of this system.
The formation of these localized structures has been recently discussed
using {\em exact} solutions by Yannouleas and Landman for two electron 
parabolic dots \cite{Yan00} and, for rings with 6 electrons by Koskinen 
{\em et al.}  \cite{Kos00}. 
These authors have computed the exact excitation spectrum and shown that 
it adjusts to rotor bands, thus revealing the existence of the electron
molecule.

Even when the exact wave function is known it is not a trivial task
to notice the existence of a Wigner molecule in the intrinsic frame. 
It can be hinted by means of intrinsic distribution functions, like the 
conditional probability density $\rho(r_2,\theta_{21}|\, r_1=\xi)$, which 
gives the probability of finding an electron 
at $(r_2, \theta_{21})$ when the other one is fixed at $r_1=\xi$. 
Therefore, it provides valuable insight on the intrinsic structure of the 
two-electron system. In Ref.\ \cite{Berxx} Berry reviews the use of the 
conditional probability density to describe the formation of electron molecules 
in the doubly excited states of the helium atom. In what follows we show 
that a similar analysis can be used for the two-electron rings.

Figure 10 shows the conditional probability density when the first 
electron is placed at its most probable value for the radial coordinate
in both V1 and V3 confining potentials. Since this position is taken as 
the origin for the angle $\theta_{21}$ this implies that the first electron is 
placed in those graphs at $(\xi,0)$, with $\xi=1.5$, 1.8~$a_0^*$ for V1 and V3, 
respectively.
Notice that the two distributions yield a peak for the second electron 
that lies opposite to the first one, i.e., at $(-\xi,0)$. 
On a first look both functions look rather similar
although a closer inspection reveals that the inner hole is more 
pronounced in the case of V3. This is in agreement with a stronger 
localization of the second electron in this case. In order to 
manifest more clearly this result it is helpful to construct the
density in the intrinsic frame and compare it with the MF one. 
To this end we assume that once the most
probable relative positions are determined,
$(x,y)=(\xi,0)$ and $(-\xi,0)$, the intrinsic frame density 
can approximately be obtained by centering in each position the conditional 
probability density. This procedure leads to the two lower
plots of Fig.\ 10. Quite interestingly, 
the V3 result is clearly breaking the circular symmetry and both 
are in excellent agreement with the mean field
densities of Fig.\ 9. 

Although qualitative the preceding discussion 
supports the interpretation that the MF describes the intrinsic 
structure of the ring. 
Consequently, the existence of a high barrier in V3 produces  
an incipient Wigner molecule. However, it remains as an interesting task 
for the future to obtain the full set of excited states and discuss 
its rotor-like character, following Refs.\ \cite{Yan00,Kos00} 
as well as the analytical expressions of Ref.\ \cite{Wen96}.

\subsection{Unrestricted  time-dependent LSDA (TDLSDA)}

Based on the mean-field picture, the TDLSDA allows to describe 
the collective dipole oscillations. These scheme has been used 
to describe both spin and density excitations in quantum dots and rings in
a symmetry-restricted approach \cite{tdlsda,Emp99b}. In Ref.\ \cite{Pue99}
we described the application of TDLSDA relaxing the constraint of circular 
symmetry and using a real-time approach similar to that of Sec.\ III.  
Within TDLSDA this implies the solution of the time-dependent Kohn-Sham 
equations, which are the analog of Eq.\ \ref{eq5} within the LSDA.
We have applied this technique to the present two-electron rings and show the 
corresponding singlet and triplet spectra for V1, V2 and V3 in Fig.\ 11. 

The TDLSDA provides a rather good description 
of the rings' dipole spectra, as compared to the exact ones. 
We notice that the overall evolution with $B$
and the location of the strength is quite well reproduced for each branch. 
However, finer details of the spectra, such as the fragmentation structures 
are not reproduced in many cases. Although based on a single-particle 
picture the unrestricted TDLSDA corrects the 
lower branch instability found in Fig.\ 7. This is a manifestation
of the interaction effects taken into account by the TDLSDA. 
From the present results we conclude that, despite the 
present rings only contain two electrons, the TDLSDA provides a rather 
sensible picture of the dipole oscillations. 
   
\section{Conclusions}

In this work we have theoretically analyzed two-electron rings 
motivated by recent measurements of their FIR absorption. We have 
found that in order to explain the experimental spectra it is necessary 
to assume that the sample contained a mixture of rings with different 
barrier heigths. As expected, high barrier rings are characterized by a 
clear violation of Kohn's theorem, showing a sizeable amount of strength 
at energies $\omega > 15$~meV for $B < 6$~T. On the other hand, 
low barrier rings approximately fulfill Kohn's theorem, with an effective 
$\omega_0$, and have lower energy excitation branches more similar to 
the measured ones. A mixture of these two types of rings has been used 
to explain the main features of the experiment. A more detailed 
analysis would require including intermediate barrier rings (ideally 
knowing the actual barrier-height distribution in the sample), for 
which the spectra lie in between the preceding two extreme situations. 
 Despite the very small number of constituents, these
systems possess a rich FIR spectrum with an intrincate $B$-dependence.  

The evolution of angular momentum and total spin with barrier height 
and magnetic field has been discussed and contrasted with independent 
particle predictions. In agreement with previous studies we find  
that interaction effects become less important for increasingly narrow 
rings, obviously reflecting a bigger electronic separation than in a 
compact structure and a corresponding increase in angular momentum.

The comparison with the symmetry-unrestricted LSDA has allowed us to discuss 
the formation of an incipient Wigner molecule for the V3 case and, in general, 
has shown the good quality of this approximation.
We have proposed a method to obtain the density in the intrinsic reference 
frame, based on the conditional probability density, that quantitatively agrees 
with the LSDA result. The FIR absorption in TDLSDA provides a good overall 
strength distribution and dispersion with magnetic field, although it 
misses some finer details of the fragmentation patterns. 

\acknowledgments

This work has been performed under Grant No.\ PB98-0124 from DGESeIC, Spain.

\begin{figure}[h]
\begin{center}
\psfig{figure=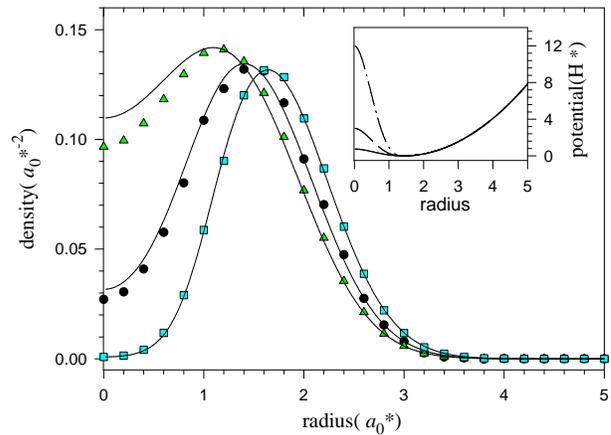,width=3.2in,clip=}
\end{center}
\caption{ Electron density for the
three rings (symbols) as a function of $r$. Notice that the exact 
solution always yields circularly symmetric densities. 
In order of decreasing density at $r=0$ the results 
correspond to V1, V2 and V3, respectively. Also shown in this 
plot are the LSDA densities (solid lines), circularly averaged in the 
case of broken symmetry (V3). Inset: External potential $V(r)$ for 
rings V1 (solid), V2 (dash) and V3 (dot-dash).
}
\end{figure}
\begin{figure}[h]
\begin{center}
\psfig{figure=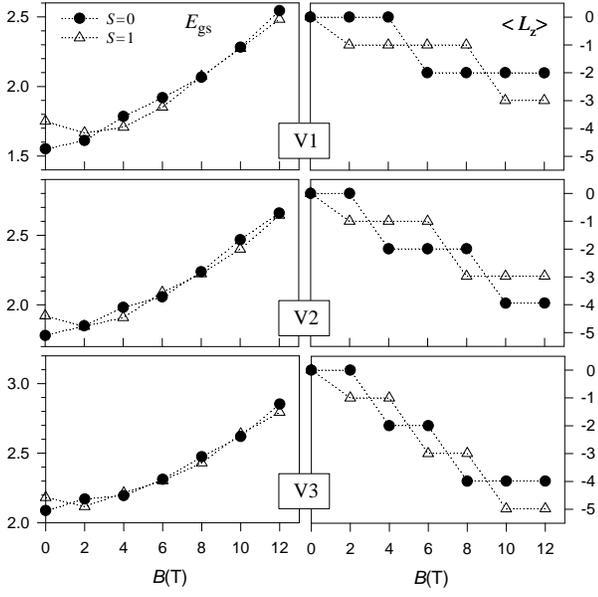,width=3.2in,clip=}
\end{center}
\caption{Evolution with magnetic field of the lowest singlet and triplet 
states for the three rings. Left panels show the state energy (in H$^*$) and 
right ones its corresponding total angular momentum in the 
perpendicular direction $\langle L_z\rangle$, in units of $\hbar$.}
\end{figure}
\begin{figure}[h]
\psfig{figure=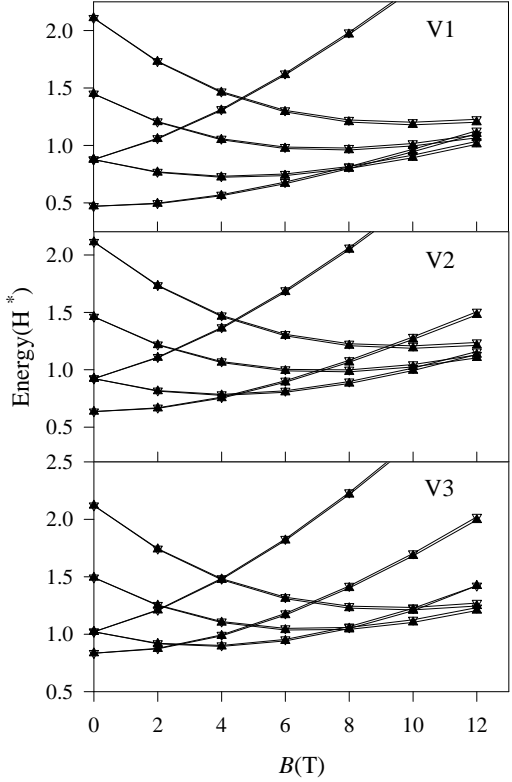,width=2.8in,clip=}
\caption{Dependence of the single particle energy levels with $B$ 
for the three potentials. The different branches, taken in vertical order at 
$B=1$T, correspond for all three cases to 
$(n \, \ell)=(1 \, 0),(1 \, -1),(1 \, 1),(1 \, -2),(1 \, -3)$, 
where $n$ (radial) and $\ell$ ($z$-component of angular momentum) 
are the usual quantum numbers. Each branch presents a small
spin splitting for increasing $B$, with the spin up states placed below the 
spin down ones.}
\end{figure}
\begin{figure}[h]
\psfig{figure=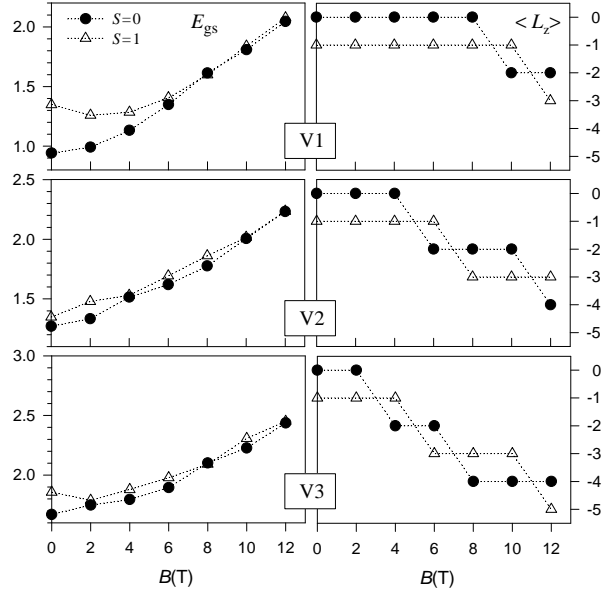,width=3.2in,clip=}
\caption{Same as Fig.\ 2 for the non-interacting case.}
\end{figure}
\begin{figure}[h]
\caption{Evolution of the FIR spectrum for singlet and triplet 
ground states in the three rings. The results for V1, V2 and V3  
are displayed in the three rows from top to bottom while left and 
right columns correspond to singlet and triplet states, respectively. 
For comparison, the dashed lines show the analytical predicition from 
Kohn's theorem (\ref{eq7b}), see Sec.\ III.A.}
\end{figure}
\begin{figure}[h]
\caption{Superposition of V1 (solid lines) and V3 (dashed) spectra at 
different magnetic fields. The experimental data are taken from Lorke 
{\em et al.} [9]. See text (Sec.\ III.A)  for details.}
\end{figure}
\begin{figure}[h]
\caption{Same as Fig.\ 5 for the non-interacting case.}
\end{figure}
\begin{figure}[h]
\psfig{figure=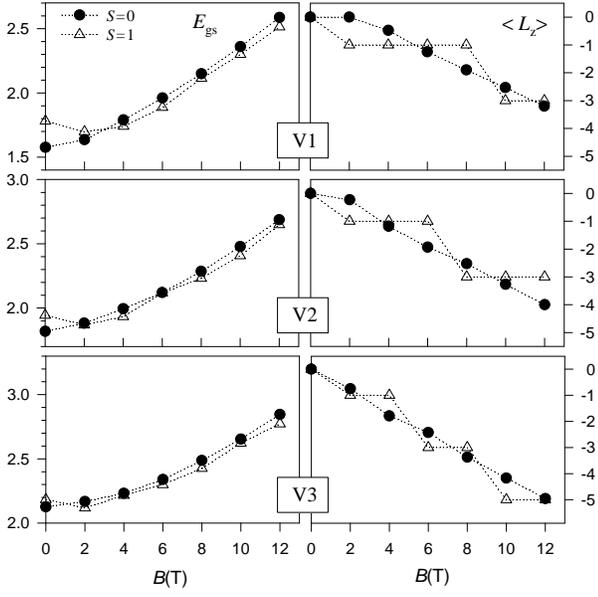,width=3.2in,clip=}
\caption{Same as Fig.\ 2 within the LSDA.}
\end{figure}
\begin{figure}[h]
\caption{Results obtained within the LSDA for $B=0$ T: 
Lower plots display the 
density $\rho$ and magnetization $m$ 
for V3 while the upper one shows the density for V1. 
In the latter case the magnetization vanishes everywhere.}
\end{figure}
\begin{figure}[h]
\caption{Upper plots: conditional probability densities (CPD) 
$\rho(r_2,\theta_{21}|\, r_1=\xi)$ obtained from the exact wave 
function at $B=0$ T.
The first electron has been fixed at $(\xi,0)$, indicated by a 
solid dot, where $\xi=1.5,1.8 \; a_0^*$ is the most probable radius 
for a single electron, 
in V1 and V3 respectively.
Lower plots: Intrinsic-frame density built from the upper 
conditional probability densities as explained in the text (Sec.\ IV.B).}
\end{figure}
\begin{figure}[h]
\caption{Same as Fig.\ 5 within the LSDA.}
\end{figure}


\begin{references}
\bibitem{units} The usual effective units system is defined by 
$\hbar=e^2/\epsilon=m=1$, where $\epsilon$ is the dielectric constant
and $m$ is the effective electron mass (given in terms of the bare mass
$m_e$ as $m=m^* m_e$). Taking the GaAs values $\epsilon=12.4$ and $m^*=0.067$
we get the effective hartree unit H$^*\approx12$~meV and effective Bohr radius
$a_0^*\approx 98$~\AA.
\bibitem{units2} For bulk GaAs the effective gyromagnetic factor is 
$g^*=-0.44$.
\end{references}
\end{document}